\documentclass{emulateapj}
\usepackage{apjfonts}

\newcommand{\ergcms}{erg cm$^{-2}$ s$^{-1}$}
\newcommand{\den}{N$_{\rm e}$}
\newcommand{\tem}{T$_{\rm e}$}
\newcommand{\cmthree}{cm$^{-3}$}
 
\newcommand{\chbeta}{C$_{H\beta}$}

\shorttitle{The Infrared Spectrum and Abundances of IC~2448}
\shortauthors{Guiles et al.}

\begin{document}

\title{The Spitzer/IRS Infrared Spectrum and Abundances of the Planetary Nebula IC~2448}

\author{S.~Guiles, J.~Bernard-Salas }
\affil{Center for Radiophysics and Space Research, Cornell
  University, 106 Space Sciences Building, Ithaca, NY 14853-6801, USA.
  sg283@cornell.edu, jbs@isc.astro.cornell.edu}

\author{S.~R.~Pottasch}
\affil{Kapteyn Astronomical Institute, P.O. Box 800, NL 9700 AV
  Groningen, the Netherlands. pottasch@astro.rug.nl}

\and
\author{T.~L.~Roellig}
\affil{NASA Ames Research Center, MS 245-6, Moffett Field, CA 94035-1000, USA. thomas.l.roellig@nasa.gov}

\begin{abstract}
  
  We present the mid-infrared spectrum of the planetary nebula
  IC~2448. In order to determine the chemical composition of the
  nebula, we use the infrared line fluxes from the {\it Spitzer}
  spectrum along with optical line fluxes from the literature and
  ultraviolet line fluxes from archival IUE spectra. We determine an
  extinction of C$_{H\beta}$ = 0.27 from hydrogen recombination lines
  and the radio to H$\beta$ ratio. Forbidden line ratios give an
  electron density of 1860 \cmthree{} and an average electron
  temperature of 12700 K. The use of infrared lines allows us to
  determine more accurate abundances than previously possible because
  abundances derived from infrared lines do not vary greatly with the
  adopted electron temperature and extinction, and additional
  ionization stages are observed. Elements left mostly unchanged by
  stellar evolution (Ar, Ne, S, and O) all have subsolar values in
  IC~2448, indicating that the progenitor star formed out of
  moderately metal deficient material.  Evidence from the {\it
    Spitzer} spectrum of IC~2448 supports previous claims that IC~2448
  is an old nebula formed from a low mass progenitor star.

\end{abstract}


\section{Introduction}


Determining accurate abundances of planetary nebulae (PNe) is
important for understanding how stars and galaxies evolve
\citep{kaler1985}. PNe abundances of elements made in low and
intermediate mass stars (such as helium, carbon, and nitrogen) can be
used to test stellar evolution models \citep{kwok2000}.  Abundances of
elements which are not changed during evolution of low and
intermediate mass stars (such as neon, argon, sulfur, and in some
cases oxygen), can give insight into the chemical content of the gas
from which the progenitor star formed \citep{kwok2000}; thus
abundances of PNe dispersed throughout a galaxy can be used to test
galactic evolution models.

IC~2448 is an elliptical, average sized PN located at
RA~=~$09^{\mathrm{h}}07^{\mathrm{m}}06^{\mathrm{s}}.26$,
DEC~=~$-69\degr56\arcmin30\arcsec.7$ (J2000.0,
\citet{kerber2003}). Optical emission line images of IC~2448 show that
diffuse [\ion{N}{2}] and [\ion{O}{3}] emission pervade the same oval
region, which agrees with IC~2448 being an old, evolved nebula
\citep{palen2002}. \citet{mccarthy1990} give further evidence for
IC~2448's advanced age, finding that its evolutionary age is 8400
years and its dynamical age is 7000 years.  IC~2448 has an H$\alpha$
diameter of $10.\arcsec7$ x $10.\arcsec0$ \citep{tylenda2003}. Thus at
a distance of 2.1 $\pm$ 0.6 kpc \citep{mellema2004}, IC~2448 has an
H$\alpha$ size of 0.11 x 0.10 pc.

Two optical surveys and one optical+ultraviolet survey of PNe include
abundance determinations for IC~2448 \citep{torres-peimbert1977,
  milingo2002b, kingsburgh1994}. Here we report the first use of
mid-infrared line fluxes from IC~2448 to determine its abundances.
Using infrared (IR) lines to derive abundances has several advantages
over using optical or ultraviolet (UV) lines \citep{rubin1988,
  pottasch1999}.  First, the correction for extinction in the IR is
smaller than in the optical and UV, and therefore errors in the
extinction coefficient and law affect IR line fluxes less.  Second, IR
lines are less sensitive to uncertainties in the electron temperature
because they come from levels close to the ground level. Finally, some
ions have lines in the IR spectrum of IC~2448, but not in the optical
or UV spectra. When combined with ionic lines of these elements
observed in the optical and UV, we have line fluxes for more ions of
these elements than previous studies, reducing the need for ionization
correction factors (ICFs) to account for unseen ionization stages.

In this paper we use the {\it Spitzer} IR spectrum supplemented
by the optical and UV spectra to derive ionic and total element
abundances of He, Ar, Ne, S, O, N, and C in IC~2448. The next section
describes the {\it Spitzer} observations and the data reduction.  \S 3
gives the optical and UV data.  In \S 4 we derive the extinction,
electron density, electron temperature, and ionic and total element
abundances.  \S 5 compares the abundances of IC~2448 with solar and
discusses the nature of the progenitor star.

\section{ Spitzer Observations and Data Reduction}
\label{spitzer_obs_s}

IC~2448 was observed with all four modules (Short-Low (SL), Long-Low
(LL), Short-High (SH), and Long-High (LH)) of the Infrared
Spectrograph (IRS) \citep{houck2004} on the {\it Spitzer Space
  Telescope} \citep{werner2004} as part of the GTO program ID 45. The
AORkeys for IC~2448 are 4112128 (SL, SH, LH observed 2004 July 18),
4112384 (SL, SH, LH off positions observed 2004 July 18), and 12409088
(LL observed 2005 February 17).  The data were taken in `staring mode'
which acquires spectra at two nod positions along each IRS slit. For the
on target AORkeys (4112128 and 12409088), the telescope was pointed at
RA~=~$09^{\mathrm{h}}07^{\mathrm{m}}06^{\mathrm{s}}.4$,
DEC~=~$-69\degr56\arcmin31\arcsec$ (J2000.0). For the off target
AORkey (4112384), the telescope was pointed at
RA~=~$09^{\mathrm{h}}07^{\mathrm{m}}06^{\mathrm{s}}.6$,
DEC~=~$-69\degr58\arcmin29\arcsec$ (J2000.0).  Peak-up imaging was
performed for AORkey 4112128, but not for the other AORkeys.

The data were processed through version s14.0 of the {\it Spitzer}
Science Center's pipeline. We begin our analysis with the
unflatfielded ({\it droopres}) images to avoid potential problems in
the flatfield. Then we run the {\it irsclean}\footnote{This program is
  available from the Spitzer Science Center's website at
  http://ssc.spitzer.caltech.edu} program to remove rogue and flagged
pixels, using a mask of rogue pixels from the same campaign as the
data. Next we remove the background.  To do this we use the off
positions for SL, SH, and LH; for LL we use the off order (for
example, LL1 nod1 - LL2 nod1).  The high resolution spectra (SH and
LH) are extracted from the images using a scripted version of the
SMART program \citep{higdon2004}, using full-slit extraction.  Due to
the extended nature of IC~2448, the low resolution spectra (SL and LL)
are extracted from the images manually in SMART using a fixed column
extraction window of width 14.0 pixels ($25.\arcsec 2$) for SL and 8.0
pixels ($40.\arcsec 8$) for LL. The spectra are calibrated by
multiplying by the relative spectral response function which is
created by dividing the template of a standard star (HR~6348 for SL
and LL, and $\xi$~Dra for SH and LH) by the spectrum of the standard
star extracted in the same way as the source (\citet{cohen2003};
G.~C.~Sloan, private communication).  Spikes due to deviant pixels
missed by the {\it irsclean} program are removed manually.

The large aperture LH (11\arcsec.1 x 22\arcsec.3) and LL (10\arcsec.5
x 168\arcsec) slits are big enough to contain all of the flux from
IC~2448. This is supported by the fact that the continuum flux from
IC~2448 in LH matches that from LL with no scaling between them.
However, the smaller aperture SH (4\arcsec.7 x 11\arcsec.3) and SL
(3\arcsec.6 x 57\arcsec) slits are too small to contain the entire
object. Thus we scale SL up to LL and SH up to LH. SL has a scaling
factor of 2.30 and SH has a scaling factor of 3.00.  No scaling factor
is needed for orders within a module (for example, SL1, SL2, and SL3
all have the same scaling factor of 2.30). Additionally, there is no
need of a scaling factor between the two nod positions.

\begin{figure}
\plotone{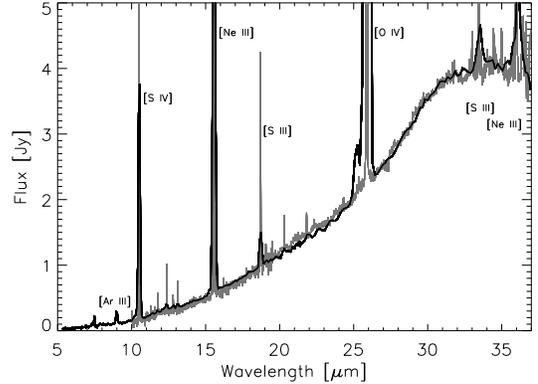}
\caption{The scaled, nod averaged, {\it Spitzer} IRS spectrum of
  IC~2448. The low resolution spectrum is shown in black and the high
  resolution spectrum in grey. SL is scaled up to LL, and SH is scaled
  up to LH because SL and SH both have small apertures that do not
  receive the entire flux from IC~2448. Some of the lines have peaks
  above 5 Jy (the strongest line, \ion{O}{4} at 25.89 \micron, peaks
  at 270 Jy in the HR spectrum), but the y-axis range is restricted in
  the plot to reveal the continuum.  \label{spectrum}}
\end{figure}

\begin{figure}
\plotone{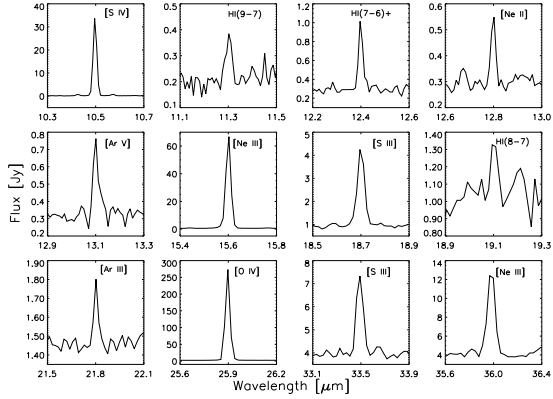}
      \caption{Close-ups of emission lines in the scaled, nod averaged
        {\it Spitzer} IRS high resolution spectrum of IC~2448. The
        \ion{H}{1}(7-6)+ line is a blend of the \ion{H}{1}(7-6) line
        (which contributes most of the flux) and the \ion{H}{1}(11-8)
        line. \label{HRlines}}
\end{figure}

Figure \ref{spectrum} shows the average of the two nods of the {\it
  Spitzer} IRS spectrum of IC~2448.  The peak of the continuum at
$\sim$30~\micron ~in F$_\lambda$ units implies that the dust is cool,
$\sim$100~K. No evidence of polycyclic aromatic hydrocarbons (PAHs) or
silicate dust is seen in the spectrum. We see lines from ions of H,
Ar, Ne, S and O in the IR spectrum of IC~2448. Close-ups of the lines
in the high resolution {\it Spitzer} IRS spectrum of IC~2448 are shown
in Figure \ref{HRlines}. High ionization lines of [\ion{O}{4}]
(ionization potential IP = 55 eV) and [\ion{Ar}{5}] (IP = 60 eV) are
observed, but even higher ionization lines such as [\ion{Ne}{5}] (IP =
97 eV) and [\ion{Mg}{5}] (IP = 109 eV) are not observed, indicating
that there is only a moderately hard radiation field. Low ionization
lines that would come from the photodissociation region (PDR) such as
[\ion{Ar}{2}] (IP = 16 eV) and [\ion{Si}{2}] (IP = 8 eV) are not
observed. A complete list of the observed lines and their fluxes is
given in Table \ref{ir_lines}.  The highest flux lines in the SH
module (10.51 \micron [\ion{S}{4}] and 15.55 \micron [\ion{Ne}{3}])
have bumps on each side of them that are instrumental artifacts,
possibly resulting from internal reflection in the SH module or an
effect of photon-responsivity. The bumps are just visible in Figure
\ref{HRlines} for the [\ion{S}{4}] line. However, the bumps contain a
negligible amount of flux -- only $\sim$3\% and $\sim$1\% of the flux
in the main [\ion{S}{4}] and [\ion{Ne}{3}] lines respectively, and we
did not include these small contributions in our line flux
measurements.

The line fluxes are measured interactively in SMART for each nod
position by performing a linear fit to the continuum on both sides of
the line and then fitting a gaussian to the line. The values of the
average line fluxes from both nod positions are given in Table
\ref{ir_lines}.  We estimate uncertainties in the line fluxes in two
ways.  In the first method, we propagate the uncertainties in the
gaussian fit to the line flux to determine the uncertainty on the
average flux from both nod positions.  In the second method, we use
the standard deviation of the fluxes measured in the two nod positions
to determine the uncertainty on the average flux from both nod
positions.  We then take the final uncertainty as the greater of these
two uncertainties.  Lines fluxes typically have uncertainties
$\lesssim$~10\%; lines with larger uncertainties ($\lesssim$~15\% and
$\lesssim$~30\%) are marked in Table \ref{ir_lines}. We determine
3$\sigma$ upper limits for lines not observed but relevant for the
abundance analysis (denoted by a less-than sign in Table
\ref{ir_lines}).  Upper limits are obtained by calculating the flux
contained in a gaussian with width determined by the instrument
resolution and height equal to three times the root mean square (RMS)
deviation in the spectrum at the wavelength of the line.

\begin{table}
\begin{center}
  \caption{{\it Spitzer} IRS observed line fluxes \label{ir_lines}}
\begin{tabular}{rrr|rrr} 
\tableline
\tableline
$\lambda$(\micron) & Line & Flux\tablenotemark{a} & $\lambda$(\micron) & Line & Flux\tablenotemark{a} \\
\tableline
    6.99  &  [\ion{Ar}{2}]        &   $<$13.9                    &  14.32 &   [\ion{Ne}{5}] &   $<$4.83  \\
    7.90  &   [\ion{Ar}{5}]       &   $<$25.7                    &  15.55 & [\ion{Ne}{3}]   &      2080  \\
    8.99  & [\ion{Ar}{3}]         &      90.7                    &  18.71 &  [\ion{S}{3}]   &      103   \\
    10.51 &   [\ion{S}{4}]        &      1780                    &  19.06 & \ion{H}{1}(8-7) &      9.67\tablenotemark{d} \\ 
    11.30 & \ion{H}{1}(9-7)       &      12.0\tablenotemark{c}   &  21.82 & [\ion{Ar}{3}]   &      6.74  \\
    12.37 & \ion{H}{1}(7-6) + \tablenotemark{b} & 30.5           &  24.30 &   [\ion{Ne}{5}] &   $<$3.68  \\ 
    12.81 &  [\ion{Ne}{2}]        &      10.7\tablenotemark{c}   &  25.89 &   [\ion{O}{4}]  &      4280  \\
    13.10 &   [\ion{Ar}{5}]       &      18.3\tablenotemark{d}   &  33.47 &  [\ion{S}{3}]   &      62.3  \\
    13.53 &   [\ion{Mg}{5}]       &   $<$4.36                    &  36.01 & [\ion{Ne}{3}]   &      154   \\

\tableline
\end{tabular}
\tablecomments{All fluxes are from the high resolution modules except
  the lines with $\lambda$ $<$ 10 \micron ~which are from the SL module.
  A less-than sign indicates a three sigma upper limit.}

\tablenotetext{a}{Observed flux in units of 10$^{-14}$ \ergcms{}. Flux
  uncertainties are $\lesssim$~10\% unless otherwise noted.}

\tablenotetext{b}{The \ion{H}{1}(7-6) and \ion{H}{1}(11-8) lines are
  blended in the spectrum.}

\tablenotetext{c}{Flux uncertainty $\lesssim$~15\%.}
\tablenotetext{d}{Flux uncertainty $\lesssim$~30\%.}
\end{center}
\end{table}

\section{Optical and UV Data}

We complement our IR line fluxes with optical and UV line fluxes in
order to determine abundances. The optical and UV data provide line
fluxes from ions not seen in the infrared spectrum (especially carbon
and nitrogen). We obtain optical line fluxes from
\citet{milingo2002a}. They observed IC~2448 with the 1.5 m telescope
and Cassegrain spectrograph at the Cerro Tololo Inter-American
Observatory in the Spring of 1997 using a 5 \arcsec x 320\arcsec
~slit. The slit width is about half of the diameter of IC~2448, and so
\citet{milingo2002a} missed some of the nebular flux. We assume that
the optical lines measured in the small aperture are representative of
the entire nebula of IC~2448 because IC~2448 has evenly distributed
optical [\ion{N}{2}] and [\ion{O}{3}] line emission \citep{palen2002}.
The extinction corrected fluxes for the lines we use are listed in
Table \ref{op_lines} as given by \citet{milingo2002a}. These authors
report uncertainties in their line fluxes of $\lesssim$~10\% for their
strong lines (with strengths $\geq$ H$\beta$) and have flagged
uncertainties of $\gtrsim$~25\% and $\gtrsim$~50\% for the weaker
lines.  These uncertainties are given in Table \ref{op_lines}.

\begin{table}
\begin{center}
  \caption{Extinction corrected optical line fluxes relative to
    H$\beta$=100 from \citet{milingo2002a} \label{op_lines}}
\begin{tabular}{rrr|rrr}
\tableline
\tableline
$\lambda$ (\AA) & Line & Flux\tablenotemark{a} & $\lambda$ (\AA) & Line & Flux\tablenotemark{a} \\
\tableline
 3727 &   [\ion{O}{2}]   & 5.1\tablenotemark{b} &  6584 &   [\ion{N}{2}]  & 1.0   \\
 3869 &   [\ion{Ne}{3}]  & 103                  &  6717 &   [\ion{S}{2}]  & 0.1   \\
 4070 &   [\ion{S}{2}]   & 1.1                  &  6731 &   [\ion{S}{2}]  & 0.1   \\
 4363 &   [\ion{O}{3}]   & 16.4                 &  7005 &   [\ion{Ar}{5}] & 0.1   \\
 4471 &   \ion{He}{1}    & 3.3                  &  7135 &   [\ion{Ar}{3}] & 4.8   \\
 4686 &   \ion{He}{2}    & 40.3                 &  7236 &   [\ion{Ar}{4}] & 0.1\tablenotemark{c}   \\
 4740 &   [\ion{Ar}{4}]  & 4.7                  &  7264 &   [\ion{Ar}{4}] & 0.1\tablenotemark{c}   \\
 4959 &   [\ion{O}{3}]   & 386                  &  7323 &   [\ion{O}{2}]  & 0.6   \\
 5007 &   [\ion{O}{3}]   & 1173                 &  7751 &   [\ion{Ar}{3}] & 1.2   \\
 5755 &   [\ion{N}{2}]   & 0.1\tablenotemark{b} &  9069 &   [\ion{S}{3}]  & 2.5   \\
 6312 &   [\ion{S}{3}]   & 0.5                  &  9532 &   [\ion{S}{3}]  & 5.2   \\
 6436 &   [\ion{Ar}{5}]  & 0.1\tablenotemark{b} &       &                 &       \\
\tableline
\end{tabular}
\tablenotetext{a}{Extinction corrected flux relative to
  H$\beta$=100. The H$\beta$ flux in the same aperture is 724.4
  $\times$ 10$^{-14}$ \ergcms{} \citep{milingo2002a}. Flux uncertainties
  are $\lesssim$~10\% unless otherwise noted.}

\tablenotetext{b}{Flux uncertainty $\gtrsim$~50\%.}
\tablenotetext{c}{Flux uncertainty $\gtrsim$~25\%.}
\end{center}
\end{table}

High and low resolution large aperture International Ultraviolet
Explorer (IUE) spectra of IC~2448 from the IUE Newly Extracted Spectra
(INES) system are available on the web\footnote{The IUE INES system
  archive website is http://ines.vilspa.esa.es}. The high resolution
spectra we use are labeled SWP19067 and LWR15096, and the low
resolution spectra we use are labeled SWP03194 and LWR02756. We use
SMART to measure the line fluxes from the spectra, and the results are
listed in Table \ref{uv_lines}. Uncertainties are obtained from the
gaussian fit to each line and are $\lesssim$~15\% unless otherwise
noted. The IUE large aperture (10\arcsec x 23\arcsec ~ellipse) is big
enough to contain essentially all of the flux from IC~2448, and no
aperture scaling factor needs to be applied to the spectra.

\begin{table}
\begin{center}
\caption{Selected observed ultraviolet line fluxes \label{uv_lines}}
\begin{tabular}{crl|crl}
\tableline
\tableline
$\lambda$ (\AA) & Line & Flux\tablenotemark{a} & $\lambda$ (\AA) & Line & Flux\tablenotemark{a} \\
\tableline
 1241 & \ion{N}{5}  &  8.86\tablenotemark{b} & 1907 &  \ion{C}{3}]  & 2250  \\  
 1483 & \ion{N}{4}] &  58.5\tablenotemark{b} & 2326 &   \ion{C}{2}] & 12.6\tablenotemark{b,c}  \\
 1548 & \ion{C}{4}  &  1900                  & 2422 & [\ion{Ne}{4}] & 184    \\
 1640 & \ion{He}{2} &  2180                  & 2471 &  [\ion{O}{2}] &  4.77\tablenotemark{b,d}  \\
 1750 & \ion{N}{3}] &  75.5                  &      &               &   \\

\tableline
\end{tabular}
\tablenotetext{a}{Observed flux in units of 10$^{-14}$ \ergcms{}. Unless
  otherwise marked, all fluxes are from IUE low resolution spectra and
  have uncertainties of $\lesssim$~15\%. }

\tablenotetext{b}{Flux from IUE high resolution spectra.}
\tablenotetext{c}{Flux uncertainty $\lesssim$~50\%.}
\tablenotetext{d}{Flux uncertainty $\lesssim$~30\%.}
\end{center}
\end{table}

\section{Data Analysis}

Our goal is to calculate element abundances in IC~2448. In order to do
this, we must first determine the extinction toward and physical
conditions within the nebula.  We iterate to find self-consistent
solutions for the electron density and temperature.  Then we use the
derived values of extinction, electron density, and electron
temperature to derive ionic abundances.  Finally we sum the observed
ionization stages of each element to derive total elemental
abundances.

  \subsection{Extinction Correction}
  \label{ext_subsxn}
  We calculate the amount of interstellar extinction in two
  ways. First, we compare the observed H$\beta$ flux for the whole
  nebula with the H$\beta$ flux predicted from infrared hydrogen
  recombination lines for case B recombination for a gas at \tem{} = 10
  000 K and \den{} = 1000 \cmthree{} using the theoretical hydrogen
  recombination line ratios from \citet{hummer1987}. The
  \ion{H}{1}(7-6) and \ion{H}{1}(11-8) are blended in the spectrum,
  and theoretically the \ion{H}{1}(11-8) line should be 12.26\% of the
  \ion{H}{1}(7-6) line \citep{hummer1987}; thus this amount is
  subtracted out before predicting the H$\beta$ flux from the
  \ion{H}{1}(7-6) line. The results are shown in Table
  \ref{ext}. Using the average predicted H$\beta$ flux from that
  table, and the total observed H$\beta$ flux of 1410 $\times$
  10$^{-14}$ \ergcms{} \citep{acker1992}, we obtain \chbeta =0.33.

  \begin{table}
  \begin{center}
  \caption{Derivation of extinction coefficient from hydrogen recombination lines \label{ext}}
  \begin{tabular}{rcrc}
  \tableline
  \tableline
  $\lambda$ (\micron) & Line  & ${F_{Line}}_{Observed}$\tablenotemark{a} & ${F_{H\beta}}_{Predicted}$\tablenotemark{b} \\
  \tableline
  11.30        & \ion{H}{1}(9-7)                     & 12.0                & 3922  \\
  12.37        & \ion{H}{1}(7-6) + \tablenotemark{c} & 30.5                & 2808  \\
  19.06        & \ion{H}{1}(8-7)                     & 9.67                & 2336  \\   
  \tableline
               & Average predicted H$\beta$ flux     &                     & 3022  \\
   \tableline
   \end{tabular}
   \tablenotetext{a}{Observed line flux in 10$^{-14}$ \ergcms{}.}
   \tablenotetext{b}{Predicted H$\beta$ flux in 10$^{-14}$ \ergcms{}.}
   \tablenotetext{c}{The \ion{H}{1}(7-6) and \ion{H}{1}(11-8) are
     blended in the spectrum. The contribution of the \ion{H}{1}(11-8)
     line is removed before predicting the H$_\beta$ flux (see \S
     \ref{ext_subsxn}).}

   \end{center}
   \end{table}
       
   The second method we use to determine the extinction is to compare
   the observed H$\beta$ flux to that predicted by the 6 cm radio
   flux using the following equation from \citet{pottasch1984}:
   \begin{displaymath}
     F(H\beta) = { S_{6 cm} \over 2.82 \times 10^9 t^{0.53} (1 + {He^+ / H^+} + 3.7{He^{++} / H^+})} 
   \end{displaymath}
   where t is the electron temperature in 10$^4$ K and $2.82 \times
   10^9$ does the unit conversion so that $S_{6 cm}$ is in Jy and
   F(H$\beta$) is in \ergcms{}. Using ionic helium abundances from \S
   \ref{abundances_sxn}, t=1.27, and $S_{6 cm}$ = 0.089 $\pm$ 0.008 Jy
   \citep{gregory1994}, we predict that F(H$\beta$) = 2345 $\times$
   10$^{-14}$ \ergcms{}, which gives \chbeta = 0.22.  This value is
   close to the value derived from the hydrogen recombination lines of
   \chbeta=0.33.
 
   We correct the IR and UV lines for extinction using the average of
   the results from our infrared hydrogen recombination line and radio
   predictions, \chbeta = 0.27 (corresponding to E$_{\bv}$ = 0.18)
   along with the extinction law from \citet{fluks1994}.  We note that
   our value of the extinction is higher than in some previous
   studies.  \citet{milingo2002a} use the ratio of H$\alpha$/H$\beta$
   to get \chbeta = 0.09.  \citet{torres-peimbert1977} use Balmer line
   ratios to get \chbeta = 0.15. Perhaps the short baselines used to
   measure extinction in those studies led to a low value of \chbeta.
   In contrast, \citet{kingsburgh1994} use radio data to predict
   H$\beta$ and get \chbeta = 0.40.  We use the extinction corrected
   optical line fluxes from \citet{milingo2002a} as opposed to
   applying our own extinction correction to the observed fluxes
   because their extinction correction gives the correct Balmer
   decrement and corrects for calibration errors.

  \subsection{Electron Density}

  We assume an electron temperature (\tem{}) of 12700 K to derive the
  electron density (\den{}). We justify this choice of \tem{} in \S
  \ref{e_temp_subsxn}; however, we use ratios of extinction corrected
  fluxes from pairs of lines of the same ion close in energy to derive
  the electron density, and these line ratios depend only slightly on
  \tem{}, so an error in the adopted \tem{} will not greatly affect the
  determination of \den{}.  The results of the density calculations are
  shown in Table \ref{e_density}.  The \ion{S}{2} density is
  unreliable because the lines used to determine it are weak (about
  1/1000 of the H$\beta$ flux) and their ratio is in the non-linear
  regime of the line ratio versus density curve.  The \ion{Ne}{3}
  15.55 \micron/36.01 \micron ~ratio is also in the non-linear regime
  and gives an imprecise density.  The \ion{Ar}{3} 8.99 \micron/21.82
  \micron ~ratio is in the low density limit.  We adopt a density of
  \den{} = 1860 \cmthree{} derived from the \ion{S}{3} line ratio as it
  provides the most accurate measurement of the density.

  There is a wide range of values for \den{} in the literature.  For
  example, \citet{torres-peimbert1977} give \den{} from the intersection
  of [\ion{O}{2}] and [\ion{N}{2}] line ratios on a plot of \tem{} vs
  \den{} as 12000 \cmthree{} and \den{} from the H$\beta$ flux as 1700
  \cmthree{}.  Whereas \citet{milingo2002b} give \den{} = 10 \cmthree{} from
  the [\ion{S}{2}] lines, which is an inaccurate density indicator in
  this case as discussed above.  Finally \citet{kingsburgh1994} give
  \den{} = 490 \cmthree{} from the [\ion{Ar}{4}] 4711 \AA/4740 \AA ~lines,
  but those lines are weak (only $\sim$ 5\% of the H$\beta$ flux), and
  the 4711 \AA ~line is blended with \ion{He}{1}, which makes the
  density derived from this line ratio unreliable. For the most part,
  abundances are not greatly affected by the adopted value of the
  density. For the range of densities for IC~2448 given in the
  literature (\den{} = 10--12000 \cmthree{}), all abundances are within
  35\% of their values for \den{} = 1860 \cmthree{}, except the sulfur
  abundance which is a factor of two larger at \den{} = 12000 \cmthree{}
  than at \den{} = 1860 \cmthree{}.

\begin{table}
\begin{center}
  \caption{Electron Densities assuming \tem{} = 12700 K \label{e_density}}
\begin{tabular}{lcrrc}
\tableline
\tableline
ion           & lines used    & Ioniz. Potential & Line Ratio & \den{}          \\ 
              &   (\micron)   &         (eV)     &            & (\cmthree{})     \\
\tableline
\ion{S}{2}    & 0.6731/0.6717 & 10.36            &  1.00     &      781       \\
\ion{S}{3}    & 18.71/33.47   & 23.33            &  1.66     &     1860        \\
\tableline
\end{tabular}
\tablecomments{The \ion{S}{3} lines provide the most accurate measurement of
  the density, and so we assume a density of 1860 \cmthree{}.}
\end{center}
\end{table}

  \subsection{Electron Temperature}
  \label{e_temp_subsxn}

  We use extinction corrected fluxes from pairs of lines of the same
  ion widely separated in energy to derive the electron temperature
  (\tem{}) assuming an electron density of \den{} = 1860 \cmthree{}.  The
  results are shown in Table \ref{e_temperature}. It is possible to
  determine \tem{} from pairs of ionic lines not in the table
  (e.g. \ion{N}{2} 5755 \AA/6584 \AA ~and \ion{Ar}{5} 7005 \AA/13.10
  \micron); however, the \tem{} determined from those ratios is not as
  accurate because one or both of the lines in each of those ratios is
  not high quality.  We adopt \tem{} = 12700 K from the average of the
  electron temperatures in the table. We use this value of the
  temperature in the subsequent abundance analysis. For comparison,
  \citet{torres-peimbert1977} give T(\ion{O}{3}) = 12500 K.
  \citet{milingo2002b} give T(\ion{O}{3}) = 12500 K,
  T(\ion{N}{2})=22400 K, and T(\ion{S}{3}) = 25200 K; however, they
  deem the values of T(\ion{N}{2}) and T(\ion{S}{3}) unrealistic.
  \citet{kingsburgh1994} give T(\ion{O}{3}) = 11000 K.  We note that
  there is no correlation of \tem{} with ionization potential, although
  such a correlation has been noticed in some previous studies
  (e.g. \citet{pottasch1999,bernard-salas2001}).

\begin{table}
\begin{center}
  \caption{Electron Temperatures assuming \den{} = 1860 \cmthree{} \label{e_temperature}}
\begin{tabular}{lcrrr}
\tableline
\tableline
ion           & lines used    & Ioniz. Potential & Line Ratio & \tem{}     \\
              & (\micron)     &       (eV)       &            & (K)       \\
\tableline
\ion{S}{3}    & 0.6312/18.71  &  23.33           & 0.125      & 13900    \\
\ion{Ar}{3}   & 0.7751/8.99   &  27.63           & 0.334      & 11500    \\
\ion{O}{3}    & 0.4363/0.5007 &  35.12           & 0.0140     & 13100    \\
\ion{Ne}{3}   & 0.3869/15.55  &  40.96           & 1.27       & 12400    \\
\tableline
\end{tabular}
\end{center}
\end{table}

  \subsection{Abundances}
\label{abundances_sxn}

\begin{table}
\begin{center}
\caption{Ionic and Elemental Abundances \label{abundances}}
\begin{tabular}{lcrrrr}
\tableline
\tableline
Ion & $\lambda$ (\micron) & N$_{\it ion}$/N$_{\it H^+}$ & N$_{\it elem}$/N$_{\it H^+}$  \\
\tableline
He$^{+}$  & 0.4471                    & 0.060               & \\
He$^{++}$ & 0.4686                    & 0.034               & 0.094 \\      
\tableline
Ar$^+$ \tablenotemark{a}   & 6.99    & $<$4.21$\times10^{-08}$ &   \\
Ar$^{++}$ & 8.99, 21.82               & 3.57$\times10^{-07}$  &   \\
Ar$^{3+}$ & 0.4740                    & 7.73$\times10^{-07}$  &   \\
Ar$^{4+}$ & 13.10                     & 1.82$\times10^{-08}$  & 1.15$\times10^{-06}$ \\
\tableline
Ne$^+$   & 12.81                     & 5.24$\times10^{-07}$  &   \\
Ne$^{++}$ & 15.55                     & 4.96$\times10^{-05}$  &   \\
Ne$^{3+}$ & 0.2422                    & 1.41$\times10^{-05}$  &   \\
Ne$^{4+}$ & 24.30                     & $<$1.31$\times10^{-08}$ & 6.42$\times10^{-05}$ \\
\tableline
S$^{++}$  & 18.71, 33.47              & 4.15$\times10^{-07}$  &   \\
S$^{3+}$  & 10.51                     & 1.53$\times10^{-06}$  & 1.95$\times10^{-06}$ \\
\tableline
O$^+$    & 0.7323                    & 2.82$\times10^{-06}$  &   \\
O$^{++}$  & 0.4363, 0.4959, 0.5007    & 2.10$\times10^{-04}$  &   \\
O$^{3+}$  & 25.89                     & 3.86$\times10^{-05}$  & 2.51$\times10^{-04}$ \\
\tableline
N$^+$    & 0.6584                    & 1.20$\times10^{-07}$  &   \\
N$^{++}$  & 0.1750                    & 2.39$\times10^{-05}$  &   \\
N$^{3+}$  & 0.1483                    & 3.09$\times10^{-05}$  &   \\
N$^{4+}$ \tablenotemark{b} & 0.1241   & 4.20$\times10^{-06}$  & 5.49$\times10^{-05}$ \\
\tableline
C$^+$    & 0.2326                    & 2.09$\times10^{-06}$  &   \\
C$^{++}$  & 0.1907                    & 1.64$\times10^{-04}$  &   \\
C$^{3+}$  & 0.1548                    & 1.05$\times10^{-04}$  & 2.71$\times10^{-04}$ \\
\tableline
\end{tabular}
\tablenotetext{a}{Ar$^+$ may originate from the PDR (instead of the
  ionized region). However, it does not affect the total argon
  abundance because the upper limit is small compared to abundances of
  other argon ions.}

\tablenotetext{b}{N$^{4+}$ probably originates from the star and thus
  it is not included in the total nitrogen abundance.}
\end{center}
\end{table}

\begin{table}
\begin{center}
  \caption{Comparison of abundances in IC~2448 to other sources \label{comp_abundances}}
\begin{tabular}{lcccccc}
\tableline
\tableline
\multicolumn{1}{c}{Element}&\multicolumn{4}{c}{IC~2448 Abundances}&\multicolumn{1}{c}{Solar\tablenotemark{d}} &\multicolumn{1}{c}{IC~2165\tablenotemark{e}}  \\
\cline{2-5}
& present  & TPP\tablenotemark{a} & M\tablenotemark{b} & KB\tablenotemark{c} & \\
\tableline
He(-2)    &  9.4    & 11    & 12  & 8.7  &  8.5 & 10  \\
Ar(-6)	  &  1.2    &  ...  & 1.2 & 0.75 &  4.2 & 1.2 \\
Ne(-5)	  &  6.4    & 8.3   & 6.8 & 11   &  12  & 5.7 \\	
S(-6)	  &  2.0    &  ...  & 9.2 & 6.8  &  14  & 4.5 \\
O(-4)	  &  2.5    & 4.1   & 3.3 & 5.3  &  4.6 & 2.5 \\
N(-5)	  &  5.5    & 3.2   & 9.3 & 24   &  6.0 & 7.3 \\
C(-4)	  &  2.7    &  ...  & ... & 8.6  &  2.5 & 4.8 \\
C+N+O(-4) &  5.8    &  ...  & ... & 16   &  7.7 & 8.0 \\
\tableline
\end{tabular}
\tablecomments{Numbers above should be multiplied by 10$^x$ where x is
  given in parenthesis in the left hand column to get abundances. For
  example, the abundance He/H in the present work is
  9.4$\times$10$^{-2}$.}

\tablenotetext{a}{\citet{torres-peimbert1977} from optical data.}
\tablenotetext{b}{\citet{milingo2002b} from optical data.}
\tablenotetext{c}{\citet{kingsburgh1994} from UV and optical data.}  
\tablenotetext{d}{Solar abundances from \citet{asplund2005} and
  \citet{feldman2003} as described in \citet{pottasch2006}.}
\tablenotetext{e}{IC~2165 abundances from \citet{pottasch2004}.}
\end{center}
\end{table}

We use the \den{} and \tem{} given above together with the IR, optical,
and UV extinction corrected line fluxes to determine the ionic
abundances for ions of Ar, Ne, S, O, N, and C. The helium ionic
abundances are calculated using emissivities in \citet{benjamin1999}.
Then we sum the ionic abundances for all expected stages of ionization
of an element to determine the abundance of that element.  The S$^+$
probably originates in the PDR because it has an IP of 10.4 eV; thus
it is not included in the total sulfur abundance.  The N$^{4+}$ 1241
\AA ~line probably originates from the star because no other ions with
such large ionization potentials (IP = 77.7 eV) are observed; thus it
is not included in the total nitrogen abundance.

The infrared spectrum enables us to observe lines of Ar$^{4+}$,
Ne$^{+}$, S$^{3+}$, and O$^{3+}$ that are missing or weak in the
combined optical and UV spectrum of IC~2448; and it additionally
allows us to place upper limits on the amount of Ar$^+$ and
Ne$^{4+}$. Observations of the S$^{3+}$ and O$^{3+}$ infrared lines
are particularly important for determining accurate elemental
abundances because those ionization stages contribute significantly to
the total sulfur and oxygen elemental abundances respectively.  The
results for ionic and total element abundances are shown in Table
\ref{abundances}.  We do not need to apply an ionization correction
factor (ICF) because we observe all ionization stages expected to have
a significant contribution to the elemental abundances derived here.

One might expect S$^{4+}$ to be present in the \ion{H}{2} region
because ions with higher ionization potentials are observed.  However,
a model of the PN Me 2-1, which has a star with similar temperature to
IC~2448's central star, indicates that the S$^{4+}$ contribution to
the total sulfur abundance is $\sim$ 15\%.  S$^{4+}$ is not important
because photons with enough energy to ionize S$^{3+}$ to S$^{4+}$ are
absorbed by the more abundant ions of other elements that have similar
ionization potentials (R. Surendiranath, private communication). We
assume that the S$^{4+}$ contribution to the total sulfur abundance is
negligible, and so we do not use an ICF for sulfur.

\section{Discussion}

In Table \ref{comp_abundances} we compare the abundances we derive
with those from previous works.  Our helium, argon and neon abundances
are all close to previous results.  For the remaining elements, we
discuss reasons why our results differ from previous ones below.  The
abundances of Ar, Ne, S, and O determined in this study should be more
accurate and precise than in previous studies because we have used
infrared lines that are less sensitive to temperature and extinction
than optical and UV lines.

The sulfur abundance determined in this study is lower than previously
reported. \citet{milingo2002b} used an ICF of 39.18 to account for
unseen S$^{3+}$, and such a large ICF leads to a large uncertainty in
their sulfur abundance result. \citet{kingsburgh1994} used an ICF of
3.13 to account for unseen S$^{2+}$ and S$^{3+}$. Therefore, while we
report a lower sulfur abundance than previous authors, it is more
accurate because we observe both S$^{2+}$ and S$^{3+}$ in the IR
spectrum and thus we do not need to use an ICF.

The oxygen abundance derived here is somewhat lower than given in
former studies. This is mainly due to our use of the O$^{3+}$ 25.89
\micron ~line to determine the O$^{3+}$ abundance, because the ICFs of
previous studies overestimated the amount of O$^{3+}$. We find that
the O$^{3+}$ abundance is only 15\% of the total oxygen abundance,
while ICFs used in previous studies assume that O$^{3+}$ contributes
between 42\% and 57\% of the total oxygen abundance.

The nitrogen and carbon abundances determined here and in previous
studies are uncertain.  The nitrogen abundances determined from the
two optical studies have large uncertainties because the optical
studies must use large ICFs (\citet{torres-peimbert1977} use 33.9 and
\citet{milingo2002b} use 1279.69) to account for unseen N$^{++}$,
N$^{3+}$, and N$^{4+}$.  This study and that of \citet{kingsburgh1994}
use UV lines for important ionization stages that dominate the element
abundances of nitrogen and carbon.  However, the ionic abundances
determined from UV lines are very sensitive to the adopted electron
temperature and extinction.  For example, lowering (raising) the
adopted \tem{} by just 1000 K leads to an increase of $\sim$70\%
(decrease of $\sim$40\%) in the derived nitrogen and carbon
abundances. Lowering (raising) the adopted \chbeta ~by 0.10 leads to
up to a $\sim$30\% decrease (increase) in the derived nitrogen and
carbon abundances. We use a higher \tem{} than \citet{kingsburgh1994}
(we use \tem{}=12700 K and they use \tem{}=11000 K), and we use a lower
extinction (we use \chbeta=0.27 whereas they use \chbeta=0.40).
Increasing the temperature and decreasing the extinction both have the
effect of lowering the abundances derived from UV lines, causing the
nitrogen and carbon element abundances derived in this study to be
less than those derived by \citet{kingsburgh1994}.  

The average of the Ar, Ne, and O abundances gives a metallicity of
$\sim$0.45 Z$_\sun$, with an uncertainty of roughly 30\%. These
elements are left relatively unchanged during stellar evolution. Thus
their low abundances imply that IC~2448's progenitor star formed from
metal deficient material, with a metallicity closer to that of the
large magellanic cloud than to that of the sun. Sulfur is also left
relatively unchanged during stellar evolution.  However, many PNe have
low sulfur abundances compared to solar, and so the sulfur abundance
is not used in the average metallicity calculation. The sulfur
abundance of IC~2448 is lower than the sulfur abundances of all 26 of
the PNe studied by \citet{pottasch2006}, which supports the idea of a
subsolar composition for IC~2448's progenitor star. The abundance of
helium derived here for IC~2448 is somewhat above solar, which implies
that some chemical processing took place within IC~2448's progenitor
star.  The nitrogen and carbon abundances are more uncertain, but
close to solar.

\citet{mendez1992} use their determinations of IC~2448's stellar
gravity (log {\it g} = 4.8) and effective temperature (T$_{eff}$ =
65000 K) to derive the current mass of IC~2448's star as 0.58
M$_\sun$. This current stellar mass corresponds to an initial stellar
mass of M$_{\rm i}$ $\approx$ 1 M$_\sun$ at a metallicity of Z =
0.5-1.0 Z$_\sun$ \citep{vassiliadis1993}. This low stellar mass, along
with the low metallicity of IC~2448's progenitor star, imply that the
first and perhaps the third dredge-up occurred.

First dredge-up would have increased the abundances of $^4$He,
$^{14}$N, and $^{13}$C, decreased the abundance of $^{12}$C, and left
the $^{16}$O abundance the same \citep{marigo2003}. Second dredge-up
(which increases $^4$He and $^{14}$N but decreases $^{12}$C, $^{13}$C,
and $^{16}$O) is only expected to occur if the initial mass of the
progenitor star is between 3 and 5 M$_\sun$ \citep{marigo2003}, and so
it is not expected to occur here.  Third dredge-up, which
significantly increases $^4$He and $^{12}$C and slightly increases the
amounts of some other elements \citep{marigo2003}, might have
occurred.  It is expected to occur in stars that have M$_{\rm i}
\gtrsim$ 1.5 M$_\sun$ at Z = Z$_\sun$, but this limit is at lower
masses for lower Z \citep{marigo1999}. Third dredge-up enriches the
amount of carbon relative to oxygen, and thus a C/O ratio greater than
one would indicate that third dredge-up occurred \citep{iben1983}.

The C/O ratio in IC~2448 derived here is 1.1, but the carbon abundance
of IC~2448 is very uncertain as discussed above. Thus we cannot
determine if the C/O ratio is really greater or less than one. The IR
spectrum of IC~2448 (Figure \ref{spectrum}) does not show PAHs which
are often observed in PNe with C/O $>$ 1, nor does it show silicates
which are often observed in PNe with C/O $<$ 1 \citep{zuckerman1986,
  bernard-salas2005}. The uncertainty in our carbon abundance and the
lack of PAHs and silicates in the IR spectrum do not allow us to
determine if IC~2448 is carbon-rich or oxygen-rich.

The abundances of IC~2448 are close to the abundances of PN IC~2165
which is a spherical nebula with a low mass ($\lesssim$ 3 M$_\sun$)
progenitor star \citep{pottasch2004}.  IC~2165 probably experienced
third dredge-up (in addition to first dredge-up) because it has C/O
$\sim$ 2 \citep{pottasch2004}. The elements not much affected by the
various dredge-up episodes (Ar, Ne, S, and O) all have subsolar values
in IC~2448 and IC~2165, which implies that the progenitor stars of
these nebula were created from metal deficient gas.

The IR continuum of IC~2448 gives a cool dust temperature of $\sim$
100 K, supporting previous studies that show IC~2448 to be an old,
evolved nebula.  Additionally, IR lines that would come from the
photodissociation region such as [\ion{Ar}{2}] and [\ion{Si}{2}] are
not observed. Perhaps this indicates that most of the
photodissociation region has been destroyed, which fits the picture of
IC~2448 being an old PN where the ionization front has gobbled up most
of the photodissociation region.

\section{Conclusions}

This is the first mid-IR spectral study of IC~2448.  The abundance of
helium is slightly above solar, indicating that some chemical
enrichment has occurred.  The high uncertainties in the nitrogen and
carbon abundances (due to their reliance on abundances determined from
UV lines which depend strongly on the electron temperature and to a
lesser extent on the extinction) make it difficult to determine how
much chemical enrichment occurred. The elements not affected much by
stellar evolution (Ar, Ne, S, and O) all have subsolar values in
IC~2448, indicating that the progenitor star formed out of somewhat
metal deficient material. Our use of infrared ionic lines which are
less sensitive to extinction and temperature, and some of which arise
from ions with no observable lines in the optical or UV, leads to a
more accurate determination of abundances for Ar, Ne, S, and O than
previously possible.  The abundances determined fit with the picture
of IC~2448 having a low mass progenitor star that underwent first and
perhaps third dredge-up. The IR continuum gives a cool dust
temperature of $\sim$ 100 K, supporting previous studies that show
IC~2448 to be an old, evolved nebula.  Additionally, lines that would
arise from the PDR are missing or weak, indicating that much of the
PDR is destroyed, consistent with IC~2448 being an old PN.

\acknowledgments This work is based in part on observations made with
the {\it Spitzer Space Telescope}, which is operated by the Jet
Propulsion Laboratory, California Institute of Technology under NASA
contract 1407. Support for this work was provided by NASA through
Contract Number 1257184 issued by JPL/Caltec. This work is also based
in part on INES data from the IUE satellite. This research made use of
the SIMBAD database, operated at CDS, Strasbourg, France. We thank
R. Surendiranath for his comments on the contribution of S$^{4+}$ to
the total sulfur abundance.



\end{document}